\documentclass[aps, pra,  a4paper, showpacs, twocolumn, english, 10pt]{revtex4-2}
\usepackage{bbm, amsthm, bm, textcomp, nicefrac, geometry,ragged2e}
\usepackage[dvipsnames]{xcolor}
\usepackage{float}
\usepackage[bbgreekl]{mathbbol}
\usepackage{graphicx, epstopdf, color, verbatim, enumitem}
\usepackage{pbox,array}
\usepackage{mathrsfs}
\usepackage{amssymb}
\usepackage{textgreek}
\usepackage{mathtools}
\usepackage{stackrel}
\usepackage[thinlines]{easytable}
\usepackage{amsmath}
\usepackage[utf8]{inputenc}
\makeatletter
\usepackage{soul}
\usepackage[caption=false]{subfig}
\usepackage{hyperref}
\hypersetup{
    colorlinks = true,
    linkcolor = Purple,
    citecolor = Purple,
    filecolor = Purple,
    urlcolor = Purple}

\usepackage[T1]{fontenc}
\usepackage[caption=false]{subfig}
\usepackage{babel}
\usepackage[linesnumbered,ruled]{algorithm2e}
\usepackage[normalem]{ulem}

\addto\captionsenglish{}

\newcommand\id{\mathbbm{1}}
\newcommand\ti{\text{i}}
\newcommand{\ket}[1]{\left| #1 \right\rangle}
\newcommand{\bra}[1]{\left\langle #1 \right|}

\newcommand{\bo}[1]{\boldsymbol{#1}}
\newcommand{\ketbra}[2]{\left| #1\right\rangle\!\left\langle#2\right|}
\newcommand{\proj}[1]{\left| #1\right\rangle\!\left\langle#1\right|}
\newcommand\LU{\overset{\text{LU}}{\simeq}}

\definecolor{brickred}{rgb}{0.8, 0.0, 0.0}

\begin{document}

\title{A modular entanglement-based quantum computer architecture}

\author{Ferran Riera-S\`abat and Wolfgang D\"ur}

\affiliation{Universit\"at Innsbruck, Institut f\"ur Theoretische Physik, Technikerstra{\ss}e 21a, Innsbruck 6020, Austria}

\date{\today}

\begin{abstract}
We propose a modular quantum computation architecture based on utilizing multipartite entanglement. Each module consists of a small-scale quantum computer comprising data, memory and entangling qubits. Entangling qubits are used to selectively couple different modules by harnessing some non-controllable, distance-dependent interaction, which is effectively controlled and enhanced via a proper adjusting of the internal state of the qubits. In this way, multipartite entangled states with different entanglement topologies can be shared between modules. These states are stored in memory qubits where they can be further processed so they can eventually be used to deterministically perform certain classes of gates or circuits between modules on demand, including parallel controlled-\textit{Z} gates with arbitrary interaction patterns, multi-qubit gates or whole Clifford circuits, depending on their entanglement structure. The usage of different kinds of multipartite entanglement rather than Bell pairs allows for more efficient and flexible coupling between modules, leading to a scalable quantum computation architecture.
\end{abstract}

\maketitle

\section{Introduction}
\label{sec.Introduction}

Quantum computers promise to tackle fundamental and practical problems in science, optimization, logistics, finances, chemistry, and material design that are not accessible with classical devices. However, a large number of qubits is required to harness the full power of quantum computers. Small-scale processors as are available now are already at the edge of outperforming classical devices \cite{PC_arute2019quantum, PC_wu2021strong, PC_bharti2022noisy, PC_bluvstein2024logical}, but due to the exponentially growing state space of quantum devices, their power is supposed to grow exponentially with system size. Scaling up quantum computers to make them applicable to real-world problems is thus a crucial, though challenging task. Modular architectures \cite{PC_Lekitsch_2017, PC_akhtar2023high, PC_jnane2022multicore, PC_wan2020ion} have been identified to be one possible solution, where small-scale processors are coupled and enabled to interact. Different ways to facilitate interactions and gates between modules have been proposed. This includes the shuttling of ions to some interaction zone \cite{PC_bluvstein2022quantum, PC_blakestad2009high, PC_bowler2012coherent, PC_wan2020ion}, or the usage of microwave links, waveguides, optical fibres or superconducting cables \cite{PC_moehring2007entanglement, PC_monroe2014large, PC_covey2023quantum, PC_zhong2021deterministic}. Auxiliary entanglement that is generated and possibly purified \cite{PC_dur_2007, PC_ferran_EPP_PRL, PC_ferran_EPP_PRA} can be utilized to perform gates between modules, which has the advantage that also probabilistic, low-fidelity couplings between modules suffice. What all proposals so far have in common is that they are based on bipartite entanglement, and use tunable interactions or actual transmission of particles, photons or phonons.  

In this article, we propose a modular quantum computation architecture with three distinct features. (1) Each module includes three types of units with different functionalities. The processing unit contains the data qubits which encode quantum information and store the state of the quantum processor. The entangling units serve as antennas or channels for coupling different modules. Additionally, memory units are used for storing and processing shared entanglement between modules, providing a resource for implementing intermodular gates on demand. (2) The second idea we develop in this article is an entangling unit which does not rely on direct control of interactions or particle exchange. Instead, the coupling is mediated by a distance-dependent interaction, which can be weak on its own but is effectively amplified by the use of multiple elementary qubit systems to implement a single logical qubit. By properly choosing their internal state, interaction strengths between modules can be selectively enhanced or diminished \cite{PC_riera2023remotely, PC_riera2023simulator, PC_riera2023hot}, thereby allowing the generation of selectable multipartite entangled states. (3) The last feature we introduce for our architecture generalises the two-qubit register scheme \cite{PC_twoqubitregister, PC_eisert2000optimal} and employs multipartite entanglement for performing intermodular gates. Multipartite entangled states of different kinds stored in the memory units are used as a resource to implement different classes of gates or whole circuits between modules on demand. This includes for instance parallel control-\textit{Z} gates with arbitrary pairwise patterns, multiqubit operations, Toffoli gates, or even the implementation of whole circuits composed of arbitrary sequences of Clifford gates in a single run. The latter is a feature of measurement-based computation \cite{PC_raussendorf2001one, PC_raussendorf2003measurement, PC_briegel2009measurement} that we bring to such distributed settings, where an entangled state of size $2n$ suffices to perform an arbitrary Clifford circuit acting on $n$ qubits.

In our architecture, the three units operate in parallel. The entangling units continuously generate entanglement between modules and transfer it to the memory units, which process, combine, and store the states to distribute arbitrary multipartite entanglement. Simultaneously, data qubits across all modules execute a quantum algorithm, consuming the entangled states stored in the memory units whenever an intermodular gate operation is required. Our approach is platform-agnostic, making it adaptable to various experimental setups. The three key features we introduce are both compatible and independent, allowing them to be integrated into a modular architecture either individually or in combination, providing flexibility in system design.

In the following, we describe our proposal in more detail. In Sec. \ref{sec.Modular.architecture.and.functionality} we introduce the general set-up and describe the processing, entangling and memory units of the modules. In Sec. \ref{sec.Entanglement.generation.between.modules} we show how to enhance interactions and generate entanglement between modules. We assume some weak, non-tunable interaction between the physical systems of different modules, and describe how using logical systems comprised of multiple qubits allows one to quadratically amplify interactions via the choice of internal states. We describe in Sec. \ref{sec.gate.induction} how different kinds of multipartite entangled states can be used to perform multiple gates or whole circuits between modules. We summarize and conclude in Sec. \ref{sec.Summary.and.conclusions}.


\section{Modular architecture and functionality}
\label{sec.Modular.architecture.and.functionality}

In a modular quantum computer, the quantum processor is divided into autonomous modules. Each module consists of a moderated size fully controllable multiqubit system, small enough such that full control within a module can be assumed. The central challenge is how to interconnect the modules to access the full computational capacity of the setting.

\begin{figure}
    \centering
    \includegraphics[width=0.99\columnwidth]{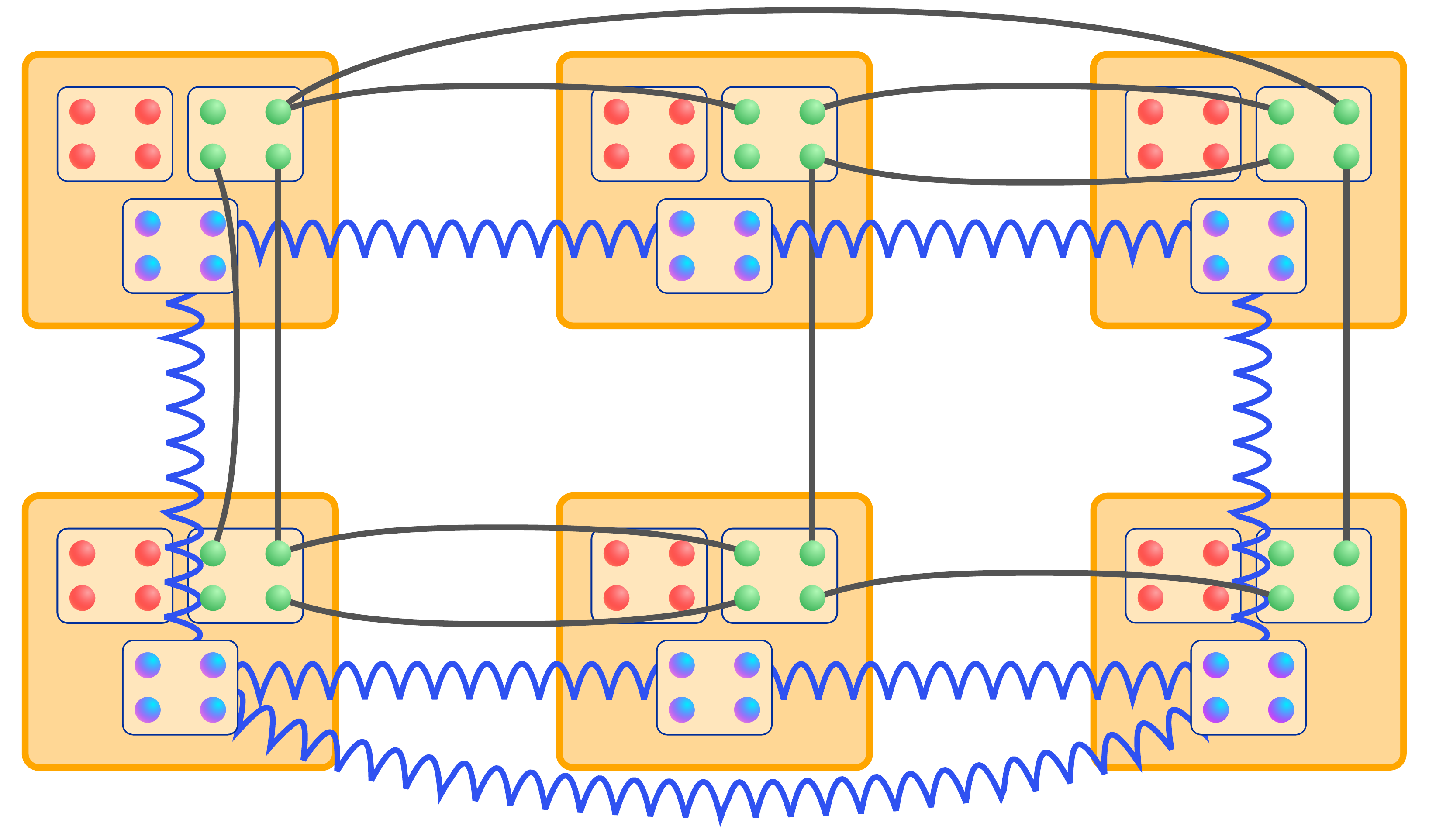}
    \caption{\label{fig.1} Each module consists of three components: the processing unit (red qubits), the memory unit (green qubits) and the entangling unit (blue qubits). The entangling units couple via a long-range physical interaction (depicted by the blue wavy lines), which influences the qubits. The interaction is amplified by encoding a logical qubit within each unit. The entangled states generated in the entangling units are then transferred to the memory units where the states are further processed and merged to prepare different kinds of entangled states with arbitrary entanglement topologies.}
\end{figure}

Our quantum computing architecture leverages entanglement to interconnect the modules, which serves as a resource for implementing intermodular operations. For this purpose, in addition to the \textit{processing unit}, each module hosts two auxiliary units. The \textit{entangling units} couple the modules with each other and are responsible for generating bipartite (or multipartite) entanglement across the modules. Meanwhile, the \textit{memory units} stores and accumulates the entangled states. Additionally, within the memory units, the entangled states can also be merged and processed, allowing the entanglement topology to extend beyond the connectivity given by the entangling units (see Fig.~\ref{fig.1}). In the following sections, we detail the functionality of each unit.

\subsection{Processing unit}

The state of the quantum processor state is encoded in the data qubits, which are distributed across different modules within the so-called \textit{processing unit}.

To operate, the quantum processor is initialized in a product state, e.g., 
\begin{equation}
    \bigotimes_{i=1}^n \ket{00\dots 0}_{\bo{D}_i}
\end{equation}
where $\bo{D}_i$ denotes the data qubits in the $i$th module, and then it is coherently evolved in a controlled manner such that different logical quantum gates are implemented to execute a certain quantum algorithm. In our architecture operations within a module are straightforward to implement. In particular, any gate of the form
\begin{equation}
    U_{\bo{D}_1\dots \bo{D}_n} = U_{\bo{D}_1}^{(1)} \otimes \cdots \otimes U_{\bo{D}_n}^{(n)}
\end{equation}
can be directly applied. However, intermodular gates are required to exploit the full computational potential of the architecture, and those cannot be directly implemented due to the separation between modules. To overcome this limitation, auxiliary multipartite entanglement shared between the modules is used to induce the different types of intermodular gates.

\subsection{Memory unit}
\label{sec.Memoryunit}

\begin{figure}
    \centering
    \includegraphics[width=\columnwidth]{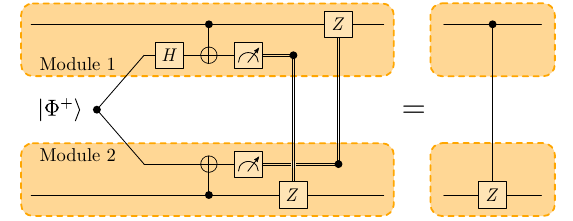}
    \caption{\label{fig:B00} Circuit representation of the implementation of an intermodular controlled-$Z$ gate using a shared Bell state, $\ket{\Phi^+} = \frac{1}{\sqrt{2}}(\ket{00} + \ket{11})$, with the assistance of classical communication.}  
\end{figure}

The memory unit serves as an auxiliary quantum processor, where entangled states (generated by the entangling unit) are stored and accumulated. Sharing entanglement between modules is crucial as, together with classical communication, they can be used to realize intermodular gates. For instance, as shown in Fig. \ref{fig:B00}, a Bell state can facilitate the execution of a control-$Z$ gate between qubits located in separate modules. Given that any multi-qubit gate can be decomposed into single-qubit operations and control-$Z$ gates \cite{PC_nielsen2002quantum}, maximizing the number of stored Bell states across all module pairs is advantageous, as it provides a mechanism for implementing multi-qubit gates on demand.

\begin{figure}
    \centering
    \includegraphics[width=0.85\columnwidth]{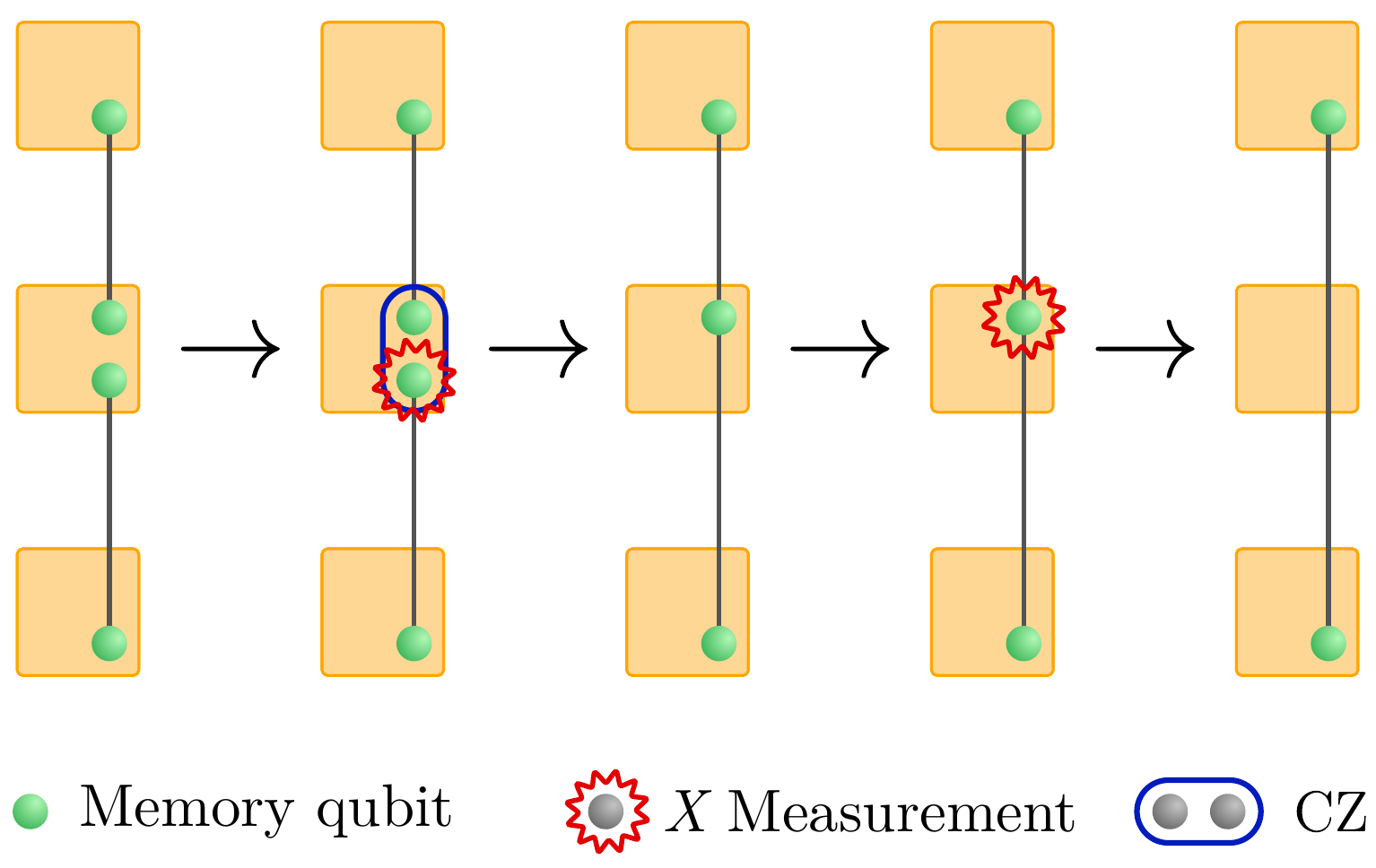}
    \caption{\label{fig:ENTANGLEMENT:SWAPING} The central module initially shares a Bell pair with each of the other modules. A control-$Z$ gate between the two qubits in the central module, followed by a projective measurement on one of the qubits establishes a GHZ state between the three modules. Measuring the other qubit in the central module, a Bell pair between the two outer modules is obtained.}  
\end{figure}

The memory unit can go beyond storing bipartite entanglement to include multipartite entangled states, such as graph states. As we show in Sec. \ref{sec.gate.induction}, a single copy of a graph state enables the implementation of the corresponding sequence of control-$Z$ gates across different modules, while a single copy of a Greenberger-Horne-Zeilinger (GHZ) state also allows for the deterministic execution of arbitrary multi-qubit $Z$-rotations, such as $e^{\ti \theta Z^{\otimes n}}$, which is a non-Clifford gate. If the entangling unit cannot directly generate the required multipartite states, the memory unit can facilitate their preparation. For instance, if the entangling units are limited to distributing bipartite entanglement, Bell states can be merged in the memory units to generate a general graph state (see Fig. \ref{fig:ENTANGLEMENT:SWAPING}). Similarly, if the entangling units lack all-to-all connectivity and can only establish direct entanglement between specific pairs of modules, the memory unit can function as a repeater node to combine states (such as through entanglement swapping \cite{PC_Briegel1998, PC_Dur1999}, see Fig. \ref{fig:ENTANGLEMENT:SWAPING}) and distribute between any pair of nodes.

This capability extends to prepare entangled states that encode various multi-qubit gates, which will eventually be applied to data qubits across different modules. It is important to note that while the deterministic implementation of intermodular gates is limited to Clifford gates (and specific non-Cliffor operations), any arbitrary quantum computation can still be systematically decomposed into a combination of Clifford and single-qubit operations. Such decompositions ensure that any quantum algorithm can be implemented, thereby enhancing the versatility of the quantum processing architecture.


\subsection{Entangling unit}
\label{sec:Entangling.unit}

The entangling unit is the component that couples to the other modules enabling the generation of entanglement between them. Various physical settings could play this role, such as optical fibres connecting the modules \cite{PC_moehring2007entanglement, PC_monroe2014large, PC_covey2023quantum} or moving atoms to induce interactions between different modules \cite{PC_Home_2009_shuttlingions, PC_blakestad2009high, PC_bowler2012coherent, PC_ruster2014experimental, PC_brown2016co, PC_bluvstein2022quantum}. In Sec.~\ref{sec:Entangling.unit}, we propose a design for the entangling units based on using physical systems inherently subjected to commuting long-range distance-dependent interactions, e.g., collective interactions induced by laser pulses \cite{PC_Porras2004, PC_richerme_2014non, PC_zhang2017observation, PC_Joshi_2020, PC_Pagano_2020} or dipole-dipole interactions \cite{PC_review-long-range-interactions, PC_Long-Distance_neutral-atoms, PC_demille2002quantum, PC_yelin2006schemes, PC_Browaeys_2016}.

As discussed in the previous subsection, distributing bipartite entanglement between modules is sufficient for implementing any intermodular gate. We also noted that shared multipartite entanglement enables the implementation of larger gates in a single step (see Sec.~\ref{sec.gate.induction}). Therefore, entangling units capable of directly generating states beyond simple bipartite entanglement would offer a significant advantage, accelerating the preparation of resource states. However, distributing multipartite entanglement is more challenging than bipartite entanglement.

The reader may note that the entangling unit would (probably) also enable the direct implementation of intermodular gates. However, we primarily use it to generate entangled states between the modules, which serve as resources for multiqubit gates. By this means, the entangling units can operate in parallel with other components. While these units sequentially produce entangled states, other components execute different parts of the computation, thereby accelerating the overall process. However, it is required that the entangling unit can generate entanglement at a faster rate than the data and memory units consume it.

\subsection{Features and possible extension}
\label{sec:addons}

The modular quantum architecture we propose offers several key capabilities that make it a powerful approach to scalable quantum computing. By leveraging a strategic division of roles within each module, we can optimize performance, enhance flexibility, and improve error resilience. Below, we discuss the main advantages and potential of this design.

A fundamental feature of our architecture is its inherent flexibility, as it is not restricted to any specific module geometry. Different arrangements of the modules can yield a variety of functionalities. For example, a larger central module can be optimized to facilitate interactions among other modules, while smaller modules may be more effective for processing quantum data or managing entanglement. Although selecting an appropriate geometry can enhance the distribution of entanglement, the entanglement topology remains largely independent of the underlying module layout. This adaptability allows the architecture to be customized for diverse computational tasks and physical implementations.

Another relevant feature of our architecture is the clear separation of roles among qubits. The memory unit in each module plays a crucial role in enhancing the overall error resilience of the system. It can be used to run entanglement purification protocols \cite{PC_dur2003multiparticle, PC_aschauer2005multiparticle, PC_Kay_2006, PC_carle2013purification}, which improve the fidelity of entangled states used for intermodular operations. This capability is particularly valuable for mitigating errors during computation, as noise is addressed before it impacts the data qubits. Additionally, the separation of roles allows for the seamless integration of non-deterministic processes into the architecture, providing greater flexibility in performing quantum gates and circuits compared to direct implementations.

In summary, our modular architecture demonstrates robust potential for scalable quantum computing. It offers enhanced interaction capabilities, improved error resilience through purification techniques, and a flexible design adaptable to various physical platforms and computational tasks.

\section{Entanglement generation between modules}
\label{sec.Entanglement.generation.between.modules}

Here we introduce a mechanism for the entangling units that does not depend on travelling particles or highly controllable interactions. Instead, it utilizes a long-range, non-tunable interaction that couples the entangling units of all modules, even if weakly.

\subsection{Two-body entangling interaction}

It is known \cite{PC_dodd2002universal, PC_benjamin2003quantum} that an $n$-qubit system subjected to two-body \textit{ZZ} interaction, i.e., a Hamiltonian of the form
\begin{equation}
    H = \sum_{1 \leq i<j \leq n} f_{ij} \, Z_i Z_j,
\end{equation}
assisted with individual control over the qubits allows one to prepare an arbitrary entangle state, i.e., any $n$-qubit state can written as
\begin{equation}
    \label{eq:arbitrarystate}
    \ket{\psi} = \prod_k \left( e^{\ti H t_k } \, \bigotimes_{i=1}^n L^{(k,i)} \right) \ket{0}^{\otimes n},
\end{equation}
where $L^{(k,i)}$ is a single qubit operation and $t_k$ an arbitrary parameter.

Therefore, if the qubits across all entangling units are coupled with each other with an interaction of this kind, any quantum circuit can be implemented and, hence, we can distribute any entangled state between the modules. However, the most natural scenario is given when the strength of the interaction decreases with the distance between the two physical systems $d_{ij}$ \cite{PC_review-long-range-interactions, PC_zhang2017observation, PC_Joshi_2020, PC_yelin2006schemes}, i.e., $f_{ij} = J /d_{ij}^{\gamma}$ where $\gamma>0$ and $J$ is the coupling constant. Such dependence on the distance impedes using such interactions for coupling the modules, as the interaction strength between two distant modules would be too weak. Nonetheless, this problem can be overcome by increasing the size of the entangling unit.

\subsection{Interaction amplification}

If the entangling unit of the $i$th module consists of $m_i$ qubits, the interaction between the $i$th and the $j$th module is given by
\begin{equation}
    \label{eq:H.ij}
    H_{ij} = \sum_{\substack{1 \leq \mu \leq m_i \\ 1 \leq \nu \leq m_j}} f_{i(\mu),j(\nu)} \, Z_{i(\mu)} \, Z_{j(\nu)}
\end{equation}
where $f_{i(\mu),j(\nu)} > 0$ is the interaction strength between the $\mu$th entangling qubit of the $i$th module and the $\nu$th entangling qubit of the $j$th module (the Latin index labels the module and the Greek index the qubit).

In each entangling unit we implement a logical qubit by restricting its state into the so-called \textit{trivial logical subspace} spanned by $\ket{\bar{0}} = \ket{0}^{\otimes m_i}$ and $\ket{\bar{1}} = \ket{1}^{\otimes m_i}$, in the logical subspace, the interaction Hamiltonian given in Eq.~\eqref{eq:H.ij} can be written as
\begin{equation}
    \label{eq:H.ij.logical}
    \bar{H}_{ij} = \sum_{k,k',l,l'=0}^1 \bra{\bar{k} \bar{l}} H_{ij} \ket{\bar{k}' \bar{l}'} \ketbra{\bar{k}\bar{l}}{\bar{k}'\bar{l}'} = \bar{f}_{ij} \, \bar{Z}_i \bar{Z}_j
\end{equation}
where $\bar{Z} = \proj{\bar{0}} - \proj{\bar{1}}$ and
\begin{equation}
    \bar{f}_{ij} = \sum_{\substack{1 \leq \mu \leq m_i \\ 1 \leq \nu \leq m_j}} f_{i(\mu),j(\nu)}
\end{equation}
is the effective coupling strength between the $i$th and the $j$th logical qubits. As we assume that the distance between physical systems within a module is much smaller than the distance between modules, we approximate $f_{i(\mu),j(\nu)} \approx J/\Delta^{\gamma}_{ij}$, where $\Delta_{ij}$ is the distance between the $i$th and the $j$th module, and then $\bar{f}_{ij} \approx m_i \, m_j J /\Delta^{\gamma}$. Note that the interaction coupling between modules increases quadratically with the number of qubits $m_i$ in each entangling unit. This enhancement can compensate for the drop in interaction strength due to the distance between the modules, enabling the establishment of strong interactions between distant modules by increasing the size of the entangling units.

For $\gamma \leq 2$, achieving all-to-all connectivity requires the size of each entangling unit to scale at most quadratically with distance. In contrast, when $\gamma > 2$, such scaling becomes impractical. For example, with dipole-dipole interactions where $\gamma = 6$, the necessary size of the entangling units rapidly becomes unfeasible. In this case, instead of pursuing all-to-all connectivity, if the entangling units can be used to establish sufficiently strong interactions between neighbouring modules the architecture can then be scaled up as needed. As we mentioned in Sec.~\ref{sec.Memoryunit}, by employing the memory units, this interaction topology enables the preparation of any multipartite entangled state.

Another important aspect is that to entangle the entangling units with each other, the logical qubit in each module must be initialized in the logical state $\ket{\bar{+}} = \frac{1}{\sqrt{2}} (\ket{\bar{0}} + \ket{\bar{1}})$, which corresponds to a GHZ state for the elementary qubits. As full control is assumed within each module, this state can be prepared using a sequence of $m-1$ control-X gates. Specifically, if $E_{i(\mu)}$ denotes the $\mu$th qubit of the $i$th entangling unit, and $\bar{E}_i$ represents the logical qubit encoded onto $E_{i(1)} \dots E_{i(m)}$, then
\begin{equation}
    \label{eq:coding}
    D_{\bar{E}_i, E_{i(1)}} \ket{+0 \dots 0}_{E_{i(1)} \dots E_{i(m)}} = \ket{+}_{\bar{E}_i},
\end{equation}
where $D_{\bar{E}_i, E_{i(1)}} = \prod_{\mu=2}^m \text{CX}_{E_{i(1)} \to E_{i(\mu)}}$ sequentially applies CX gates to entangle the elementary qubits.

Since the required initial state itself is (locally) entangled, the process is sensitive to preparation errors. It is therefore crucial to minimise these errors to ensure high-fidelity entanglement generation. Small errors during preparation can propagate through the subsequent operations, significantly affecting the quality of the resulting entanglement. However, as explained in Sec.~\ref{sec:addons}, another possibility is to utilize the memory units to reduce the noise by purifying the generated entanglement.

\subsection{Interactions control}

The trivial subspace is not the only one that simplifies $H_{ij}$ to a logical \textit{ZZ} interaction, i.e., if we encode a logical qubit in the entangling unit of the $i$th module as $\ket{\bar{0}} = \ket{\bo{k}_i}$ and $\ket{\bar{1}} = X^{\otimes m_i} \ket{\bo{k}_i}$ where $\ket{\bo{k}_i}$ is a state of the computational basis given by $Z_{i(\mu)} \ket{\bo{k}_i} = s_{i(\mu)} \ket{\bo{k}_i}$ and $s_{i(\mu)} = 1-2k_{i(\mu)}$ the Hamiltonian is also simplified to Eq.~\eqref{eq:H.ij.logical}. However, in this case, the coupling strength is given by $\bar{f}_{ij} = \sum_{\mu,\nu} f_{i(\mu),j(\nu)} s_{i(\mu)} s_{j(\nu)}$. In Ref. \cite{PC_riera2023simulator} we show how by a suitable choice of the logical subspace of each module, any interaction pattern between the logical systems can be established. However, using a different subspace leads to a reduction of the coupling strength between the logical qubits. This reference also includes an analysis of how the coupling strength varies with different interaction patterns, which depends on the geometry of the architecture. For that reason, for each interaction pattern, the maximum interaction strength needs to be evaluated. In any case, any interaction pattern can be built using the trivial subspace in a ``bang-bang'' approach, where two interacting logical qubits are encoded in the trivial subspace while the others are encoded in an insensitive subspace, see Ref. \cite{PC_riera2023simulator}. In addition, the trivial encoding improves the fidelity interaction in the presence of thermal noise. As we show in Ref.~\cite{PC_riera2023hot}, by a suitable choice of the logical subspace, the fidelity can be further enhanced at the price of a weaker coupling.

Moreover, by encoding a logical qubit in each entangling unit, we make them insensitive to the interactions between its constituents, meaning that the state of the logical qubit is not affected by the interaction of the physical qubits encoding it. This is because any logical subspace of this kind defines a decoherence-free subspace of a \textit{ZZ}-type interaction, i.e., the internal interactions of the $i$th module are described by $H_i = \sum_{\mu < \nu} f_{i(\mu),i(\nu)} Z_{i(\mu)} Z_{i(\nu)}$ and on a logical subspace it simplifies to $\bar{H}_i \propto  \bar{\id}$.

Therefore, once a logical qubit is encoded in each entangling unit, the Hamiltonian of the whole system is given by
\begin{equation}
    \label{eq:logical.hamiltonian}
    \bar{H} = \sum_{1 \leq i < j \leq n} \bar{f}_{ij} \, \bar{Z}_i \bar{Z}_j,
\end{equation}
where $\left\{ \bar{f}_{ij} \right\}$ can be tuned by a proper choice of the logical subspaces $\{ \bo{k}_i \}$. The interaction pattern is described with a graph $\mathcal{G}$ given by a set of $n$ vertices representing the logical qubits and a set of edges $\{ (i,j) \}$ between the nodes representing the interaction strength $\bar{f}_{ij}$.

\subsection{State storage}

As discussed at the beginning of this section, the Hamiltonian given in Eq.~\eqref{eq:H.ij.logical} enables us to establish entanglement between the logical qubits within the entangling units. Once a desired entangled state is prepared, it can be transferred to the memory units for further processing, merging with other entangled states, or storage. This transfer can be achieved by performing in each module a SWAP gate between the logical qubit in the entangling unit and an elementary qubit in the memory unit. However, since this operation involves both a logical qubit and an elementary qubit, it is not straightforward to implement directly. It must be decomposed into a sequence of gates acting on the individual elementary qubits within both the entangling and memory units.

The first step involves, in each module, localizing the state of the entangling unit onto a single elementary qubit. This corresponds to invert Eq.~\eqref{eq:coding}, which is achieved by applying the same gate sequence, i.e.,
\begin{equation}
    \label{eq:decoding} 
    \bigotimes_{i=1}^n D_{\bar{E}_i, E_{i(1)}} \!\! : \, \ket{\psi}_{\bar{E}_1\dots \bar{E}_n} \, \mapsto \, \ket{\psi}_{E_{1(1)}\dots E_{n(1)}}.
\end{equation}
An alternative circuit for Eq.~\eqref{eq:decoding} can be constructed using measurements combined with feed-forward operations. In this approach, each two-qubit control operation is replaced by a measurement on the target qubit, followed by a conditional operation on the control. These dynamic circuits are known to offer advantages on certain platforms \cite{PC_dynamic1, PC_dynamic2}.

\begin{figure}
    \centering
    \includegraphics[width=\columnwidth]{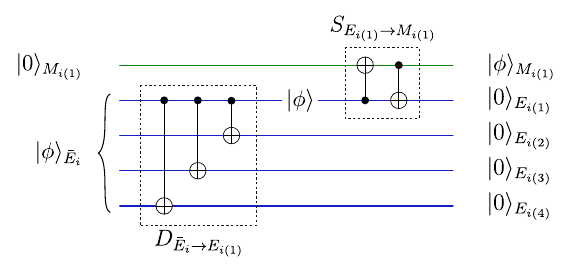}
    \caption{\label{fig:swap2}Circuit representation of Eqs.~\eqref{eq:decoding} and \eqref{eq:SWAP} which (for the trivial encoding) transfers the state of the logical qubit to a single memory qubit $M_{i(1)}$.}
\end{figure}

Then the entangled state can be transferred to the memory units by performing two control-$X$ gates per module. If we denote the $\mu$th qubit in the $i$th memory unit as $M_{i(\mu)}$, the transfer operation is given by
\begin{equation}
\begin{aligned}
    \label{eq:SWAP} 
    \bigotimes_{i=1}^n S_{E_{i(1)} \to M_{i(1)}} \!\! : \ket{\psi}_{E_{1(1)}\dots E_{n(1)}} \! \ket{0 \dots 0}_{M_{1(\mu)}\dots M_{n(\mu)}} & \\
    \mapsto \ket{\psi}_{M_{1(\mu)}\dots M_{n(\mu)}} &,
\end{aligned}
\end{equation}
where $S_{A \to B} = \text{CX}_{B\to A} \text{CX}_{A\to B}$. Following these two steps, the logical state is deterministically transferred to the memory units through a sequence of $m+1$ control-$X$ gates, involving both entangling and memory qubits. Figure \ref{fig:swap2} illustrates the full circuit for this process.

\section{Intermodular gate induction}
\label{sec.gate.induction}

In the previous section, we demonstrated how using the entangling units we can control an entangling Hamiltonian and prepare arbitrary entangled states between the different modules. The main idea of our architecture is to encode multiqubit gates in multipartite states shared by the modules. These states are used in a later stage to locally perform the corresponding gate to the data qubits of the different modules \cite{PC_wan2019quantum}. 

Before proceeding with this section, we clarify a change in notation to simplify the equations. From this point onward, when referring to the elementary qubits in the data or memory qubits, we will omit the indices that specify the qubit within each module, i.e., $D_{i(\mu)} \to D_i$ and $M_{i(\mu)} \to M_i$. This change is made because the following equations only involve a single qubit from each unit. However, there is one exception where two qubits in the same unit need to be distinguished. Then we will use the notation $M_{i(1)} \to M_i$ and $M_{i(2)} \to M'_i$.

\subsection{Diagonal gates}

First, we analyse the implementation of diagonal gates. A diagonal $n$-qubit gate $\Lambda$ can be stored in an $n$-qubit quantum register by preparing its \textit{gate state}
\begin{equation}
    \ket{\Lambda} = \Lambda \ket{+}^{\otimes n}.
\end{equation}
If we distribute the gate state $\ket{\Lambda}$ between the memory qubits in different modules, by applying local routine $T$, we can induce $\Lambda$ up to a random byproduct of local $X$ gates to the data qubits of the respective modules. $T$ consists in applying a multilateral control gate between the memory and the data qubits followed by a projective measurement on the memory units. If the measurement outcome on the $i$th memory qubit is given by $k_i \in \{ 0,1\}$, $T$ implements
\begin{equation}
    T^{(\bo{k})}_{\textit{MD}} \ket{\Lambda}_M \ket{\psi}_D
    \; \mapsto \; X^{\otimes \bo{k}} \, \Lambda \, X^{\otimes \bo{k}} \ket{\psi}_D,
\end{equation}
to the data qubits, where $M = M_1\dots M_n$, $D = D_1\dots D_n$, $T^{(\bo{k})}_{\textit{MD}} = O^{(\bo{k})}_M \bigotimes_{i=1}^n \text{CX}_{D_i \to M_i}$, $X^{\otimes \bo{k}} = X^{k_1} \otimes \cdots \otimes X^{k_n}$ and $O^{(\bo{k})} = \proj{\bo{k}}$. Notice all measurement outcomes are equally probable meaning each possible value of $\bo{k}$ is given with probability $p = 2^{-n}$, see datils in Appendix. \ref{Appendix:routine_T}. 

Therefore, $T$ only implements $\Lambda$ if $\bo{k} = \bo{0}$ is obtained. However, for the so-called Clifford gates, $\Lambda$ can always be implemented by performing a local correction operation. On the other hand, for arbitrary gates, the routine can be used as an elementary step to build a (quasi-)deterministic protocol to induce the gate. At this point, we analyse the implementation of both kinds of gates in detail.

\subsubsection{Diagonal Clifford gates}

A gate is called Clifford if it normalizes the Pauli group, i.e.,do this change becouse we never will write conisder two qubits in the same  $U$ is Clifford iff $U \, \textbf{P}_n \, U^\dagger = \textbf{P}_n$ where $\textbf{P}_n = \left\{ e^{\ti k \frac{\pi}{2}} X^{\otimes \bo{i}} Z^{\otimes \bo{j}} \, | \, \bo{i}, \bo{j} \in \{0,1\}^n, k \in \mathbbm{N} \right\}$ is the $n$-Pauli group. Therefore, if $\Lambda$ is Clifford, the outcome gate of routine $T$ is locally unitary (LU) equivalent to $\Lambda$, and it can be converted into it by applying a local correcting operation given by $C^{(\bo{k})} = \Lambda \, X^{\otimes \bo{k}} \,\Lambda^\dagger X^{\otimes \bo{k}} \in \textbf{P}_n$. Note that in this case, $\Lambda$ can be implemented deterministically by consuming a single copy of its gate state.

\textit{Pairwise two-qubit Clifford $Z$-rotation}. One example of a diagonal Clifford gate is the control-\textit{Z} gate, which is given by $\text{CZ} = e^{\ti \pi \proj{11}} \LU e^{\ti \frac{\pi}{4} ZZ}$. Given a graph $\mathcal{G}$, where a vetice represents a qubit and an edge $(i,j)$ an application of $e^{\ti \frac{\pi}{4} ZZ}$ between the $i$th and the $j$th qubits, the corresponding multiqubit gate is given by
\begin{equation}
    G = e^{\ti \frac{\pi}{4} \sum_{(k,l)\in \mathcal{G}} Z_k Z_l}.
\end{equation}
The gate state of $G$ can be directly prepared by evolving the entangling units under the interaction pattern $\mathcal{G}$. Then from $\ket{G}$, $G$ can be induced by applying $T$ followed by the correction operation $C^{(\bo{k})} = \prod_{(i,j)\in \mathcal{G}} (Z_i Z_j)^{k_i + k_j}$, see Fig.~\ref{fig:5a}.

This is a powerful class of gates as they suffice to implement any intermodular non-Clifford gate, such as multiqubit \textit{Z}-rotations, i.e., $e^{\ti \theta Z^{\otimes n}} \LU G^\dagger_{1 \dots n} \, L^{(\theta)}_1 \, G_{1 \dots n}$, where here $G_{1 \dots n} = e^{\ti \frac{\pi}{4} Z_1 (Z_2+\cdots + Z_n)}$ and $L^{(\theta)} = e^{\ti (n-1)\frac{\pi}{4}Z} e^{\ti \theta X} e^{- \ti (n-1) \frac{\pi}{4} Z}$.

\textit{Multiqubit Z-rotation}. A multiqubit $Z$-rotation
$R^{\bo{j}}\!\left( \theta \right) = e^{\ti \theta Z^{\otimes \bo{j}}}$ is Clifford for $\theta \in \left\{ \pm k \pi/4 \right\}_{k \in \mathbbm{N}}$, in particular, $R^{\bo{j}}\!\left( \pm k \pi/4 \right) \in \left\{ \id, \, Z^{\otimes \bo{j}}, R^{\bo{j}}\!\left( \pm \pi/4 \right) \right\}$. $R^{\bo{j}}(\pm \pi/4)$ is an entangling gate, and it can be induced from its gate state by applying $T$ followed by the correction operation $C^{(\bo{k})} = \left( Z^{\otimes \bo{j}}\right)^{\bo{k}\cdot \bo{j}}$ where $\bo{k}\cdot \bo{j} = \sum_{i}^n k_i \, j_i$.

\subsubsection{Diagonal non-Clifford gates}
\label{sec:Diagonal.non-Clifford.gates}

Non-Clifford gates cannot be deterministically induced from a single copy of its gate state. If the output $\bo{k} \neq \bo{0}$ is obtiained, the implemented gate $X^{\otimes \bo{k}} \Lambda X^{\otimes \bo{k}}$ is not LU to $\Lambda$. In this case, we need to use other methods. ($i$) One possibility is to decompose $\Lambda$ into a finite sequence of arbitrary multiqubit \textit{Z}-rotations. As we show later, each of these rotations can be deterministically induced by consuming a GHZ state. ($ii$) Alternatively, one can build a quasi-deterministic protocol by iterating routine $T$ where in each step the target gate is chosen in a heraldic way. We illustrate both methods with specific examples.

\textit{Two-qubit Z-rotation}. ($i$) An arbitrary two-qubit \textit{Z}-rotation can be implemented deterministically by consuming a single copy of a two-qubit maximally entangled state, i.e.,
\begin{equation}
\begin{aligned}
\label{eq:twoqubit_rotation_CZ}
    (ZZ)_{D_1D_2}^{(l)} \, Q^{(l)}_{M_1 D_1} \, Z_{M_1}^{k} \, T^{(k)}_{M_2 D_2} \ket{\text{CZ}}_{M_1M_2} \ket{\psi}_{D_1 D_2} & \\
    \mapsto e^{\ti \theta ZZ} \ket{\psi}_{D_1 D_2} &,
\end{aligned}
\end{equation}
where $Q^{(l)}_{ij} = \proj{l}_i e^{\ti \theta X} \text{CZ}_{ij}$. Note that this method consumes one ebit of auxiliary entanglement. However, in Ref.~\cite{PC_smallentanglement}, it was shown that for small angles, $e^{\ti \theta ZZ}$ can be induced by consuming less than one ebit.

($ii$) If we prepare $\ket{e^{\ti \theta ZZ}}$ and apply $T$, we implement
\begin{equation}
    T^{(\bo{k})}_{MD} \ket{e^{\ti \theta ZZ}}_M \ket{\psi}_D \mapsto e^{\ti (-1)^{k_1 + k_2} \theta ZZ} \ket{\psi}_D,
\end{equation}
where here $M = M_1M_2$ and $D = D_1D_2$. Note, with probability $p=1/2$, $e^{\ti \theta ZZ}$ is applied and the protocol is over. If not, the rotation is reversed, and we implement $e^{-\ti \theta ZZ}$, which can not be locally transformed into the target gate. In case of a failed implementation, we can perform $T$ again with $e^{\ti 2\theta ZZ}$ as the new target gate. If we succeed the overall gate is given by $e^{\ti \theta ZZ}$ and the protocol is over, but in case of failure $e^{-\ti 3\theta ZZ}$ is implemented. In case of failure, this step is iterated again, being $e^{\ti 2^{j-1}\theta ZZ}$ the target gate in the $j$th round. As each iteration succeeds with probability $p=1/2$ the protocol provides a quasi-deterministic way of implementing $e^{\ti \theta ZZ}$ where the expected number of steps required is given by $\sum_{k=1}^{\infty} k \, 2^{-k} = 2$. In addition, note that if $\theta = \pi/2^N$ where $N\in \mathbbm{N}$, in the $(N-1)$th iteration the target gate is given by $e^{\ti \frac{\pi}{4} ZZ}$ which is Clifford and therefore, it can be implemented deterministically. In Appendix. \ref{Appendix.Entanglement.cost}, we show that for $|\theta| < 0.0715 \pi$, less than one ebit of entanglement is used on average, and therefore for these values of $\theta$ this second method is more efficient.

\textit{Pairwise two-qubit \textit{Z}-rotations}. The same procedure ($ii$) can be used to implement a sequence of non-Clifford pairwise rotations, i.e., $G_\theta = e^{\ti \sum_{(k,l)\in\mathcal{G}} \theta_{kl} Z_k Z_l}$. If we prepare its gate state and apply $T$ we implement
\begin{equation}
    T^{(\bo{k})}_{\textit{MD}} \ket{G_\theta}_M \ket{\psi}_D \mapsto e^{\ti \theta \sum_{(i,j)\in \mathcal{G}} (-1)^{k_i + k_j}  Z_i Z_j} \ket{\psi}_D.
\end{equation}
For pairs $(i,j)$ such that $k_i \oplus k_j = 0$ the desired rotation is implemented, while for the other pairs, the rotations are reversed. In the next step, we iterate the procedure with $G'_{2\theta}$ as the target gate, where the graph $\mathcal{G}' = \mathcal{G} \, \backslash \, \{ (i,j) \, | \, k_i \oplus k_j = 0\}$ is given by all edges where the wrong angle was implemented. In this way, $G_\theta$ can be quasi-deterministically implemented. See an example in Appendix. \ref{appenidx:thetaG}.

\textit{Multiqubit Z-rotation}. A multiqubit $Z$-rotation $R^{\bo{j}}(\theta)$ can be implemented with a direct extension of the two methods for the two-qubit $Z$-rotation. 

\begin{figure}
    \centering
    \subfloat[]{\includegraphics[width=\columnwidth]{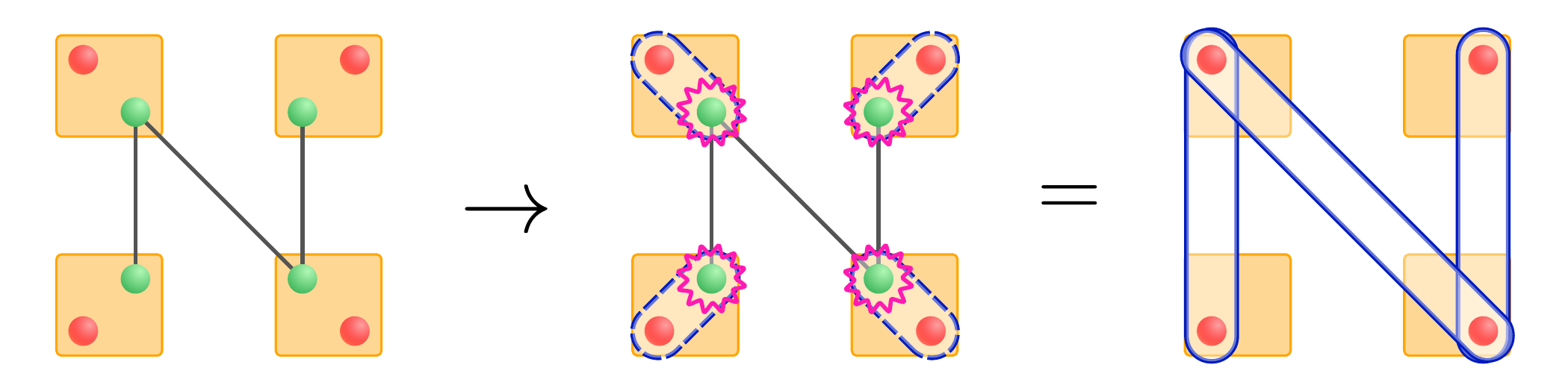}\label{fig:5a}} \\
    \subfloat[]{\includegraphics[width=\columnwidth]{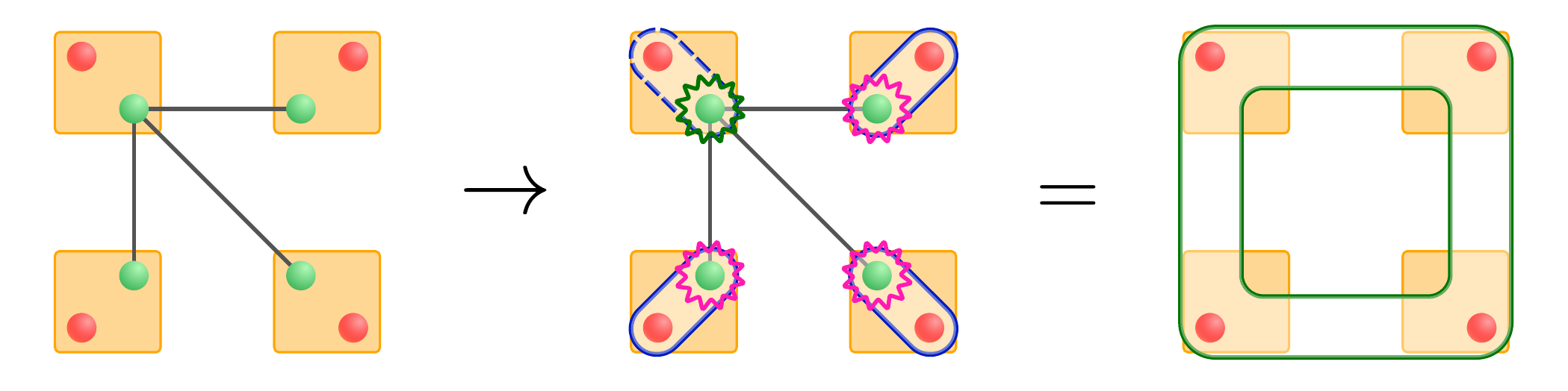}\label{fig:5b}} \\
    \subfloat[\empty]{\includegraphics[width=\columnwidth]{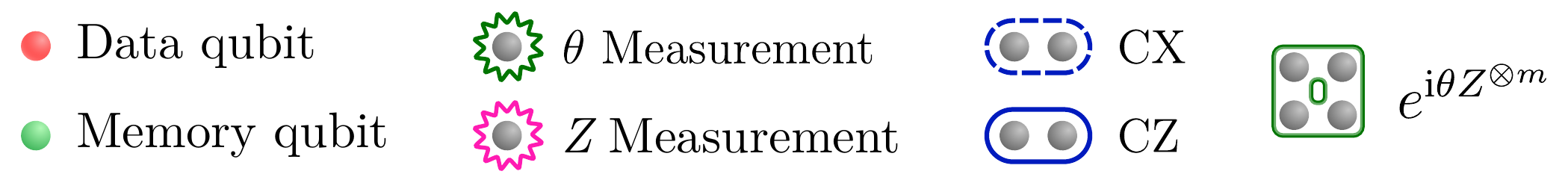}}
    \caption{(a) Implametation of a sequence of intermodular control-$Z$ gates. The data qubits (red) are in an arbitrary state, and the memory qubits (green) are initialized in the gate state which corresponds to a graph state. The routine consists in applying a CX gate between the memory and the data qubits followed by a local measurement of the memory qubits. Finally, a correction operation is performed on the data qubits. (b) Implementation of a multiqubit $Z$-rotation with an auxiliary GHZ state. First, a control gate is implemented between the memory and the data qubits. Then all memory qubits except for the central one are measured on the $Z$ basis, and a correction operation is performed on the central memory qubit. Finally, the central memory qubit is measured on the $\theta$ basis $\{ e^{-\ti \theta X} \ket{k}\}$, followed by a correction operation on all data qubits.}
\end{figure}

($i$) Equation~\eqref{eq:twoqubit_rotation_CZ} can be generalized to implement an arbitrary multiqubit $Z$-rotation by consuming a single copy of a GHZ state, i.e.,
\begin{equation}
\begin{aligned}
    (Z^{\otimes n})_D^l \, Q^{(l)}_{M_1 D_1} \, Z^{k_2+\dots+k_n}_{M_1} \, T^{(k_2,\dots,k_n)}_{M_2 D_2 \dots M_n D_n} \ket{\text{GHZ}}_M \ket{\psi}_D & \\
    \mapsto e^{\ti \theta Z^{\otimes n}} \ket{\psi}_D &
\end{aligned}
\end{equation}
where $\ket{\text{GHZ}} = \prod_{j=2}^n \text{CZ}_{1j} \ket{+}^{\otimes n}$, (see Fig.~\ref{fig:5b} and Appendix. \ref{Appendix:Multi-qubitZ_GHZ} for details).

$(ii)$ Preparing $\ket{R^{\bo{j}}(\theta)}$ and implementing $T$ we obtain an analogous situation to the two-qubit case, i.e,
\begin{equation}
    T^{(\bo{k})}_{\textit{MD}} \ket{R^{\bo{j}}(\theta)}_M \ket{\psi}_D \mapsto R^{\bo{j}} \! \left[ (-1)^{\bo{k}\cdot \bo{j}} \theta \right] \ket{\psi}_D,
\end{equation}
with probability $p=1/2$, we succed and with probability $p=1/2$ the rotation is reversed. Therefore, we can implement $R^{\bo{j}}(\theta)$ by iterating $T$ in a heraldic way analogously to the two-qubits case.

In this case, the entanglement cost of the two approaches cannot be directly compared. However, in our architecture the second method would be more suitable as the given interaction Hamiltonian allows us to directly establish GHZ states between the modules. Nevertheless, in other settings with a different set-up to interconnect the modules, it could be easier to prepare states of the form $\ket{R^{\bo{j}}(\theta)}$ than a GHZ state. In that case, the second method would be more efficient.

\textit{Toffoli gate}. An $n$-qubit non-Cilfford gate of particular interest is the Toffoli gate, as it is the elementary tool for preparing hypergraph states \cite{PC_hypergraph2013}. The Toffoli gate is given by
\begin{equation}
    \text{Toff} \, \LU e^{\ti \pi \proj{\bo{0}}}= \prod_{\bo{j} \in \{0,1\}^n} R^{\bo{j}}\!\left( \pi/2^n \right).
\end{equation}

($i$) The Toffoli gate can be implemented by sequentially performing $R^{\bo{j}}(\pi/2^k)$ for every subset of $k$ qubits, which requires to distribute of a GHZ state between each subset of qubits. Because we need to perform $(n-1)$ two-qubit control gates to prepare an $n$-qubit GHZ state, this method requires a total of $1+2^{n-1}(n-2)$ two-qubit intermodular gates. In particular, any diagonal gate $\Lambda$ can be factorized into an arbitrary multi-qubit $Z$-rotation for each subset of qubits (see Appendix. \ref{Appendix:Diagonal.gates}) and therefore it can be implemented with at most the same number of two-qubit control gates.

($ii$) The implementation following procedure is analogous to the previous examples. We prepare the gate state, apply $T$, and depending on the outcome, for some subsets of qubits the rotation is successfully implemented while for others it is reversed. Then the routine is iterated where the target gate for the $j$th step is given by a rotation of $R^{\bo{j}}\!\left(2^{j-1} \, \pi/ 2^{n}\right)$ for all sets $\bo{j}$ where the implementation failed in the previous steps. Note that in this way, at most $n-1$ steps are required as the target gate for the $(n-1)$th corresponds to a product of Clifford rotations, i.e., $\theta = \pi/4$, and therefore it can be deterministically implemented, see Appendix \ref{Appendix:toffoli} for details. 

\subsection{Non-diagonal gates}

An arbitrary $n$-qubit gate can also be stored in the memory units by preparing its Choi state, which is given by  
\begin{equation}
    \ket{\Phi_U}_{\textit{MM}'} = \frac{1}{\sqrt{2^n}} \sum_{\bo{k}\in\{0,1\}^n} \ket{\bo{k}}_M \otimes U \ket{\bo{k}}_M',
\end{equation}
where $M'_i$ and $M_i$ are two different qubits of the memory of the $i$th module. Note how in this case a $2n$-qubit quantum register where each memory unit hosts two qubits is needed.

Analogously to the diagonal case, given the Choi state of a certain gate $U$, one can find a routine $\widetilde{T}$ that induces $U$ up to a byproduct of local Pauli operators, i.e.,
\begin{equation}
\label{eq.Teleportation.U}
    \widetilde{T}^{(\bo{ij})}_{\textit{MM}'\!D} \ket{\Phi_U}_{\textit{MM}'} \ket{\psi}_D \mapsto U X^{\otimes \bo{i}} Z^{\otimes \bo{j}} \ket{\psi}_D,
\end{equation}
with probability $p=4^{-n}$ where
\begin{equation}
    \widetilde{T}^{(\bo{ij})}_{\textit{MM}'\!D} = O^{(\bo{i})}_M \, O^{(\bo{j})}_{M'} \, \text{H}^{\otimes n}_{M'} \, m\text{CX}_{M' \to M} \, \text{SWAP}_{M' D}.
\end{equation}

From Eq.~\eqref{eq.Teleportation.U}, it is straightforward to note that Clifford gates can be deterministically induced by performing a local correction operation $C^{(\bo{i},\bo{j})} = U Z^{\otimes \bo{j}} X^{\otimes \bo{i}} \, U^\dagger$. On the other hand, non-Clifford gates can not be corrected with a local gate. Instead one would need to iterate the routine where on the $j$th step the target gate is given by $U^{(j)} = U F^{(j) \dagger}$ where $F^{(j)}$ is the overall gate implemented in the previous steps where the routine failed. In this way, one can construct a (quasi)-deterministic routine for non-Clifford gates.

In this way, given a whole quantum circuit, it always is split into pieces \cite{PC_universal}, i.e., 
\begin{equation}
    U = \prod_k \left( U^{(k)} \bigotimes_{i=1}^n L^{(k)} \right),
\end{equation}
where $U^{(k)}$ are intermodular gates that can be induced with a few entangled states, e.g., Clifford gates, arbitrary multiqubit $Z$-rotations or the Toffoli gate, and $L^{(k)}$ are arbitrary modular gates. Once the circuit $U$ is factorized, then intermodular parts $U^{(k)}$ can be run in parallel and stored in the memory units of the modules. Eventually, the whole circuit can be implemented by inducing $U^{(k)}$ while intercalating the modular gates $L^{(k)}$.

\section{Summary and conclusions}
\label{sec.Summary.and.conclusions}

In this article, we introduced a modular quantum architecture designed to improve the scalability of quantum information processing. This architecture is built on three complementary, independent features: modules composed of three units, multipartite entanglement for implementing multiqubit gates, and distance-dependent interactions for generating entanglement between modules.

We demonstrated how a non-tunable two-body interaction can establish strong connections between modules without requiring direct tuning or the physical transmission of particles. A key element of this approach is the use of memory units to merge, process, and accumulate entanglement, storing intermodular quantum gates encoded in multipartite entangled states. This mechanism enables on-demand implementation of intermodular gates and facilitates the decomposition of quantum circuits into components that can be executed in parallel.

This architecture offers significant potential due to its flexibility and its ability to operate on distributed entanglement, taking advantage of well-established techniques from quantum communication. It presents an efficient and scalable pathway to quantum computation.

Future research should focus on implementing this architecture in specific physical setups, analyzing performance metrics such as error rates, and exploring how memory units can enhance intermodular gates compared to employing direct entangling units. This will be crucial for assessing the practical viability and optimization of our proposed framework.

\begin{acknowledgments}
This research was funded in whole or in part by the Austrian Science Fund (FWF) 10.55776/P36009, 10.55776/P36010 and 10.55776/COE1. For open access purposes, the author has applied a CC BY public copyright license to any author accepted manuscript version arising from this submission. Finanziert von der Europ\"aischen Union - NextGenerationEU.

We thank Pavel Sekatski for interesting discussions.

\end{acknowledgments}

\bibliography{Modular_Qcomputer.bib}

\appendix

\onecolumngrid

\section{Diagonal gates}
\label{Appendix:Diagonal.gates}

Note that a projector on a state of the computational basis can be written as
\begin{equation}
    \label{eq:projectorZZ}
    \proj{\bo{i}} = \frac{1}{2^n} \sum_{\bo{j}\in \{0,1\}^n} (-1)^{\bo{i} \cdot \bo{j}} \, Z^{\otimes \bo{j}}.
\end{equation}
Using Eq.~\eqref{eq:projectorZZ}, we can factorize any diagonal gate $\Lambda$ into a sequence of multiqubit $Z$-rotations, i.e., given an arbitrary digonal gate where $\bra{\bo{i}} \Lambda \ket{\bo{i}} = e^{\ti \alpha_{\bo{i}}}$ then
\begin{equation}
    \Lambda = \sum_{\bo{i}\in \{0,1\}^n} e^{\ti \alpha_{\bo{i}}} \proj{\bo{i}} = \prod_{\bo{i}\in \{0,1\}^n} e^{\ti \alpha_{\bo{i}} \proj{\bo{i}} } = \prod_{\bo{i},\bo{j} \in \{0,1\}^n} e^{\ti \alpha_{\bo{i}} (-1)^{\bo{i}\cdot\bo{j}} 2^{-n} Z^{\otimes \bo{j}} } = \prod_{\bo{j}\in \{0,1\}^n} R^{\bo{j}} (\theta_{\bo{j}}),
\end{equation}
where $R^{\bo{j}}(\theta) = e^{\ti \theta Z^{\otimes \bo{j}}}$ and $\theta_{\bo{j}} = \sum_{\bo{i}} (-1)^{\bo{i}\cdot \bo{j}} \alpha_{\bo{i}} / 2^n$.

Being $\textbf{P}_n = \left\{ e^{\ti \frac{\pi}{2}k} X^{\otimes \bo{x}} Z^{\otimes \bo{y}} | k \in \mathbbm{N}, \bo{x},\bo{y} \in \{0,1\}^n \right\}$ the $n$-Pauli group, $\Lambda$ is said to be Clifford iff
\begin{equation}
\begin{gathered}
    \Lambda^\dagger \, X^{\otimes \bo{k}} \, Z^{\otimes \bo{l}} \, \Lambda \in \textbf{P}_n \;\; \forall \, \bo{k}, \bo{l} \\ 
    \Leftrightarrow \; \Lambda^\dagger \, X^{\otimes \bo{k}} \, \Lambda \, X^{\otimes \bo{k}} = \Lambda^\dagger \prod_{\bo{j}} R^{\bo{j}} \! \left[(-1)^{\bo{j}\cdot\bo{k}} \theta_{\bo{j}} \right]  = \prod_{\bo{j}} R^{\bo{j}} \! \left(\omega_{\bo{jk}} \right) = \prod_{\bo{j}} \left[ \cos(\omega_{\bo{jk}}) \, \id + \ti \sin(\omega_{\bo{jk}}) \, Z^{\otimes \bo{j}} \right]  \in \textbf{P}_n \;\; \forall \, \bo{k}
\end{gathered}
\end{equation}
where $\omega_{\bo{ij}} = \left[ (-1)^{\bo{i}\cdot\bo{j}} - 1 \right] \theta_{\bo{j}} \in \{0,2\theta_{\bo{j}} \}$. Therefore, $\Lambda$ is Clifford iff it can be factorized into multiqubit rotations of the from $\theta_{\bo{j}} \in \left\{ \pm k \pi/4 \right\}_{k \in \mathbbm{N}} \;  \forall \, \bo{j}$.

\section{Induction of \textit{Z}-diagonal gates}
\label{Appendix:routine_T}

Given an arbitrary $n$-qubit diagonal gate
\begin{equation}
    \Lambda = \sum_{\bo{i} \in \{ 0,1\}^n } e^{\ti \alpha_{\bo{i}}} \proj{\bo{i}},
\end{equation}
consider its gate state $\ket{\Lambda}$ and an arbitrary $n$-qubit state
\begin{equation}
    \ket{\psi}_D = \sum_{\bo{j}} \psi_{\bo{j}} \ket{\bo{j}}_D.
\end{equation}
If we apply a multilateral control-$X$ between the two-state and measure the $M$ qubits, i.e.,
\begin{equation}
    T_{\textit{MD}}^{(\bo{k})} = \left( O^{(k_i)}_{M_1} \, \text{CX}_{D_i M_i}\right),
\end{equation}
we implement $\Lambda$ up to a byproduct of Pauli $X$ gates to the $D$ system, i.e.,
\begin{equation}
\begin{aligned}
    T_{\textit{MD}}^{(\bo{k})} \, \ket{\Lambda}_M \ket{\psi}_D = \frac{1}{\sqrt{2^n}} \, T_{\textit{MD}}^{(\bo{k})}\sum_{\bo{i},\bo{j} } e^{\ti \alpha_{\bo{j}}} \, \psi_{\bo{j}} \ket{\bo{j}}_M \ket{\bo{i}}_D = \frac{1}{\sqrt{2^n}} \, O_M^{(\bo{k})} \sum_{\bo{i},\bo{j} } e^{\ti \alpha_{\bo{j}}} \psi_{\bo{i}} \ket{\bo{j}\oplus \bo{i}}_M \ket{\bo{i}}_D & \\
    \mapsto \sum_{\bo{i}} e^{\ti \alpha_{\bo{i} \oplus \bo{k}}} \, \psi_{\bo{i}} \ket{\bo{i}}_D = X^{\otimes \bo{k}} \sum_{\bo{i}} e^{\ti \alpha_{\bo{i}}} \, \psi_{\bo{i}\oplus \bo{k}} \ket{\bo{i}}_D = X^{\otimes \bo{k}} \, \Lambda \, X^{\otimes \bo{k}} \ket{\psi}_D &,
\end{aligned}
\end{equation}
see Fig.~\ref{fig.circuit_T} for the circuit representation. Note the probability of obtaining outcome $O^{(\bo{k})}$ is independent of $\bo{k}$, i.e.,
\begin{equation}
    p(\bo{k}) = \sum_{\bo{l}} \left| \bra{\bo{k}, \bo{l}} m\text{CX} \ket{\Lambda,\psi} \right|^2 = \sum_{\bo{l}} \Big| \bra{\bo{k}, \bo{l}} \frac{1}{\sqrt{2^n}} \, \sum_{\bo{i},\bo{j} } e^{\ti \alpha_{\bo{j}}} \, \psi_{\bo{i}} \ket{\bo{j}\oplus \bo{i}, \bo{i}} \Big|^2 = \frac{1}{2^n} \sum_{\bo{l}} \left| e^{\ti \alpha_{\bo{k}\oplus \bo{l}}} \, \psi_{\bo{l}}  \right|^2 = \frac{1}{2^n} \sum_{\bo{l}} \left| \psi_{\bo{l}}  \right|^2 = \frac{1}{2^n}.
\end{equation}

\section{Induction of arbitrary gates}

Conisder an arbitrary $n$-qubit gate,
\begin{equation}
    U = \sum_{\bo{l},\bo{k}} u_{\bo{lk}} \ketbra{\bo{l}}{\bo{k}},
\end{equation}
and its Choi state
\begin{equation}
    \ket{\Phi_U}_{\textit{MM}'} = \frac{1}{\sqrt{2^n}} \sum_{\bo{k}} \ket{\bo{k}}_M \otimes U \ket{\bo{k}}_{M'} = \frac{1}{\sqrt{2^n}} \sum_{\bo{k},\bo{l}} u_{\bo{lk}} \ket{\bo{k}}_M \otimes \ket{\bo{l}}_{M'}.
\end{equation}
Given an arbitrary $n$-qubit state $\ket{\psi}_D$, we can induce $U$ up to a random byproduct of Pauli gates by applying a local routine $\widetilde{T}$ that consists in applying a multilateral control-$X$ gate between $D$ and $M$ followed by a projective measurement on the Choi state. In particular with probability $p= 4^{-n}$ we implement 
\begin{equation}
    \widetilde{T}^{(\bo{nm})}_{\textit{MM}'\!D} = O^{(\bo{n})}_M \, O^{(\bo{m})}_{M'} \, \text{H}^{\otimes n}_{M'} \, m\text{CX}_{M' \to M} \, \text{SWAP}_{M' D}.
\end{equation}
Routine $\widetilde{T}$ induces $U$ up to a byproduct of local Pauli gates, i.e.,
\begin{equation}
\begin{aligned}
    \widetilde{T}^{(\bo{rs})}_{\textit{MM}'\!D} \ket{\Phi_U}_{\textit{MM}'} \ket{\psi}_D \\
    = \frac{1}{\sqrt{2^n}} \, O^{(\bo{r})}_M \, O^{(\bo{s})}_{M'} \, \text{H}^{\otimes n}_{M'} \, m\text{CX}_{M' \to M} \sum_{\bo{k},\bo{l},\bo{j}} \psi_{\bo{j}} \, u_{\bo{lk}} \ket{\bo{k},\bo{j}}_{\textit{MM}'} \ket{\bo{l}}_D & \\
    = \frac{1}{\sqrt{2^n}} \, O^{(\bo{r})}_M \, O^{(\bo{s})}_{M'} \, \text{H}^{\otimes n}_{M'} \, \sum_{\bo{k},\bo{l},\bo{j}} \psi_{\bo{j}} \, u_{\bo{lk}} \ket{\bo{k}\oplus \bo{j}, \bo{j}}_{\textit{MM}'} \ket{\bo{l}}_D & \\
    = \frac{1}{\sqrt{2^n}} \, O^{(\bo{r})}_M \, O^{(\bo{s})}_{M'} \, \sum_{\bo{k},\bo{l},\bo{i},\bo{j}} (-1)^{\bo{i} \cdot \bo{\bo{j}}} \, \psi_{\bo{j}} \, u_{\bo{lk}} \ket{\bo{k} \oplus \bo{j}, \bo{i}}_{\textit{MM}'} \ket{\bo{l}}_D & \\
    \mapsto \sum_{\bo{k},\bo{l}} (-1)^{\bo{s} \cdot (\bo{r}\oplus\bo{k})} \, \psi_{\bo{r}\oplus\bo{k}} \, u_{\bo{lk}} \ket{\bo{l}}_D & \\    
    = U X^{\otimes \bo{r}} Z^{\otimes \bo{s}} \ket{\psi}_D &.
\end{aligned}
\end{equation}
where we have used that $H^{\otimes n} = \sum_{\bo{i},\bo{j}} (-1)^{\bo{i} \cdot \bo{j}} \ketbra{\bo{i}}{\bo{j}}$. See in Fig.~\ref{fig.circuit_WT} the circuit representation.

\section{Multi-qubit \textit{Z}-rotation}
\label{Appendix:Multi-qubitZ_GHZ}

In this section, we show in detail how a GHZ can be used to implement a multiqubit $Z$-rotation, shown in Sec.~\ref{sec:Diagonal.non-Clifford.gates}. The protocol consists of applying
\begin{equation}
    Z_{M_1}^{k_2 + \cdots + k_n} \, T^{(k_2,\dots,k_n)}_{M_2 D_2 \dots M_n D_n} = Z_{M_1}^{k_2 + \cdots + k_n} \bigotimes_{i=2}^n \left( O^{(k_i)} \text{CX}_{D_i \to M_1} \right),
\end{equation}
followed by
\begin{equation}
    (Z^{\otimes n})^l_D \, Q_{M_1 D_1} = (Z^{\otimes n})^l_D \, O^{(l)}_{M_1} \, e^{\ti \theta X_{M_1}} \text{CZ}_{M_1 D_1}, 
\end{equation}
where $O^{(k)} = \proj{k}$ to an arbitrary $n$-qubit gate $\ket{\psi}_D$ and a $\ket{\text{GHZ}}_M = \prod_{i=2}^n \text{CZ}_{1,i} \ket{+}^{\otimes n}$, i.e.,
\begin{equation}
\begin{aligned}
    (Z^{\otimes n})^l_D \, Q_{M_1 D_1} \, Z_{M_1}^{k_2 + \cdots + k_n} \, T^{(k_2,\dots,k_n)}_{M_2 D_2 \dots M_n D_n} \ket{\text{GHZ}}_M \ket{\psi}_D & \\
    = \frac{1}{2^{n/2}} \, (Z^{\otimes n})^l_D \, Q_{M_1 D_1} \, Z_{M_1}^{k_2 + \cdots + k_n} \, T^{(k_2,\dots,k_n)}_{M_2 D_2 \dots M_n D_n} \sum_{\bo{i},\bo{j}} \psi_{\bo{i}} \left(-1\right)^{j_1 (j_2+ \dots + j_n)} \ket{\bo{j}}_M \ket{\bo{i}}_D & \\
    \mapsto \frac{1}{\sqrt{2}} (Z^{\otimes n})^l_D \, Q_{M_1 D_1} \, Z_{M_1}^{k_2 +\dots + k_n} \sum_{\bo{i},j_1} \psi_{\bo{i}} \left(-1\right)^{j_1 (i_2 + k_2 + \dots + i_n + k_n)} \ket{j_1}_{M_1} \ket{\bo{i}}_D & \\
    = \frac{1}{\sqrt{2}} (Z^{\otimes n})^l_D \, Q_{M_1 D_1} \, \sum_{\bo{i},j_1} \psi_{\bo{i}} \left(-1\right)^{j_1 (i_2 + \dots + i_n)} \ket{j_1}_{M_1} \ket{\bo{i}}_D & \\
    \mapsto (Z^{\otimes n})^l_D \, \sum_{\bo{i}} \psi_{\bo{i}} \left(-1\right)^{l \sum_{k} i_k} e^{\ti (-1)^{\sum_{k} i_k} \theta}  \ket{\bo{i}}_D & \\
    = \sum_{\bo{i}} \psi_{\bo{i}} \, e^{\ti (-1)^{\sum_{k} i_k} \theta} \ket{\bo{i}}_D & \\
    = e^{\ti \theta Z^{\otimes n}} \ket{\psi}_D & ,
\end{aligned}
\end{equation}
see Fig.~\ref{fig.circuit_QT} for the circuit representation.

\section{Paiwise two-quit \textit{Z}-rotations: Particular example}
\label{appenidx:thetaG}

Conisder four modules and the target intermodular gate
\begin{equation}
    G_{\pi/16} = e^{\ti \frac{\pi}{16} (Z_1 Z_3 + Z_1 Z_4 + Z_2 Z_4 + Z_3 Z_4)}.
\end{equation}
Such a gate is non-Clifford and therefore, we need to sequentially apply routine $T$ until we succeed.

\textbf{1st step}. We prepare its gate state and apply $T$. We assume outomce $\bo{k}_1 = (0,1,0,1)$ is obtained. Then the implemented gate is given by 
\begin{equation}
    X_2 \, X_4 \, G_{\pi/16} \, X_2 \, X_4 = e^{\ti \frac{\pi}{16} (Z_1 Z_3 - Z_1 Z_4 + Z_2 Z_4 - Z_3 Z_4)}.
\end{equation}
Note the rotation is successfully implemented for pairs (1,3) and (2,4), while for pairs (1,4) and (3,4) it is reversed.

\textbf{2nd step}. We want to correct the rotation for the pairs where we failed in the last step. Therefore, in this step, our target gate is given by
\begin{equation}
    G'_{\pi/8} = e^{\ti \frac{\pi}{8} ( Z_1 Z_4 + Z_3 Z_4)}
\end{equation}
We prepare its gate state and apply $T$. If we assume outcome $\bo{k}_2 = (0, \varnothing ,1,1)$ is obtained, the implemented gate is given by
\begin{equation}
    X_3 \, X_4 \, G'_{\pi/8} \, X_3 \, X_4  = e^{\ti \frac{\pi}{8} ( - Z_1 Z_4 + Z_3 Z_4)},
\end{equation}
and the overall gate implemented is given by
\begin{equation}
    X_3 \, X_4 \, G'_{\pi/8} \, X_3 \, X_4 \, X_2 \, X_4 \, G_{\pi/16} \, X_2 \, X_4 = e^{\ti \frac{3\pi}{8} Z_1 Z_4} e^{\ti \frac{\pi}{16} (Z_1 Z_3 + Z_2 Z_4 + Z_3 Z_4)} .
\end{equation}
Note that after this second step, our initial target rotation is successfully implemented for pairs (1,3), (2,4) and (3,4), while for pair (1,4) we failed in both steps.

\begin{figure}
    \centering
    \subfloat[]{\includegraphics[width=0.13\columnwidth]{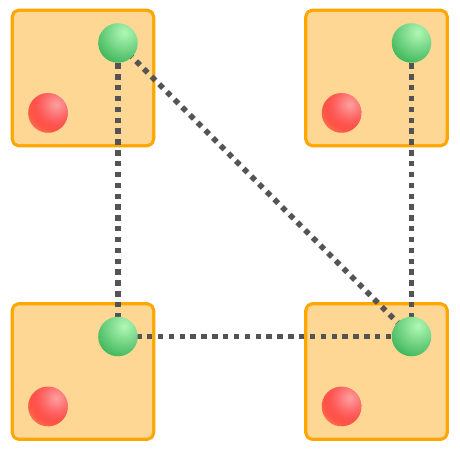}} \hfill
    \subfloat[]{\includegraphics[width=0.13\columnwidth]{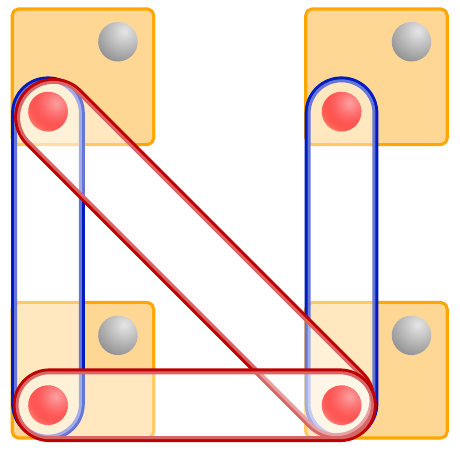}} \hfill
    \subfloat[]{\includegraphics[width=0.13\columnwidth]{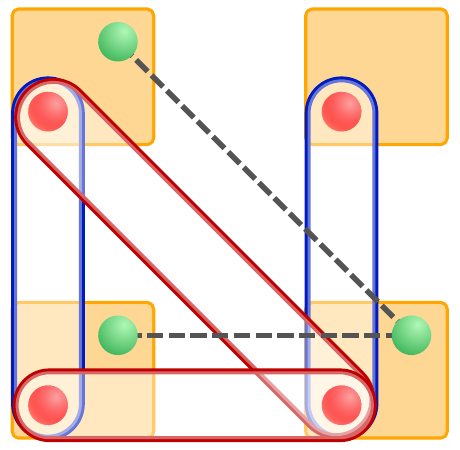}} \hfill   \subfloat[]{\includegraphics[width=0.13\columnwidth]{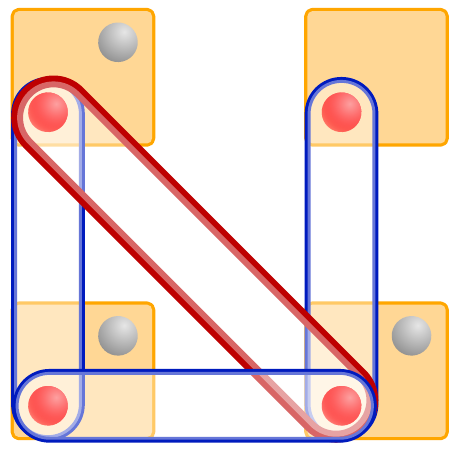}} \hfill
    \subfloat[]{\includegraphics[width=0.13\columnwidth]{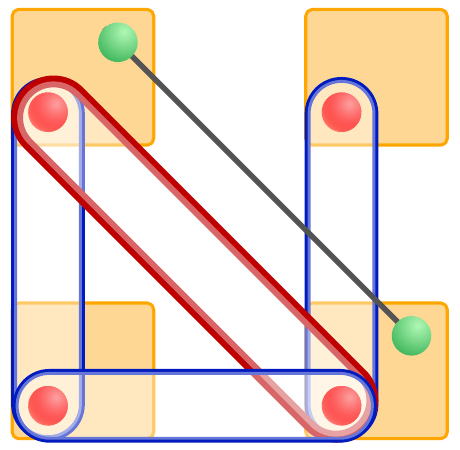}} \hfill
    \subfloat[]{\includegraphics[width=0.13\columnwidth]{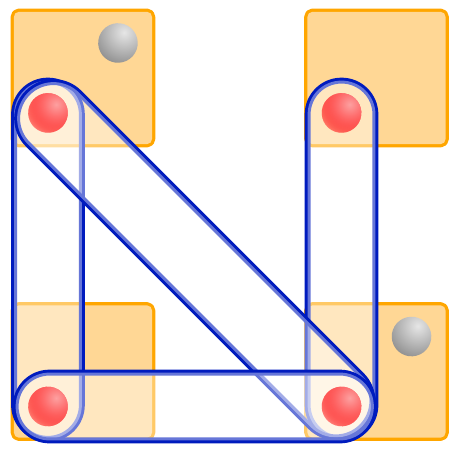}} \hfill \\
    \subfloat[\empty]{\includegraphics[width=0.5\columnwidth]{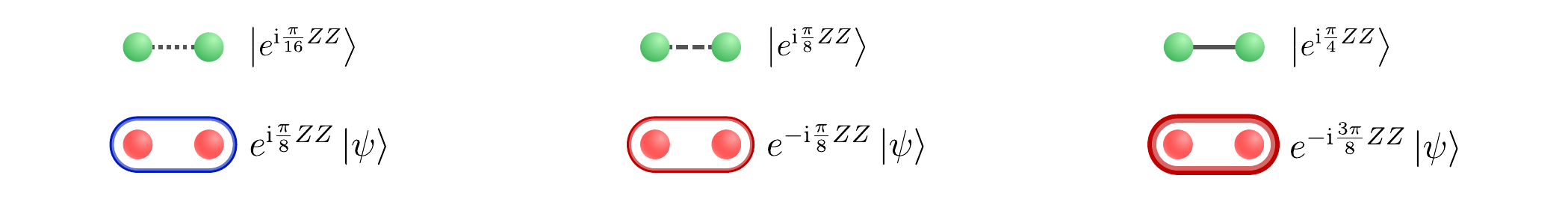}}
    
    \caption{\label{fig:loko} Implementation of $G_{\pi/16}$ where $\mathcal{G} = \{ V= \{1,2,3,4\}, E =\{ (1,3), (1,4), (2,4), (3.4)\} \}$, with routine $T$. First, we prepare the gate state $\ket{G_{\pi/16}}$ and apply $T$. If we obtain output $\bo{k}_1 = (0,1,0,1)$, the gate is only implemented correctly between qubits (1,3) and (2,4). In a second step, we prepare the gate state $\ket{G'_{\pi/8}}$ where $\mathcal{G}' = \{ V= \{1,2,4\}, E =\{ (1,4), (3.4)\} \}$ and we apply $T$. If in the second step, we obtain $\bo{k}_{2} = (0,1,1)$ we succeed in edge (3,4). In the last step we prepare the gate state $\ket{G''_{\pi/4}}$ where $\mathcal{G}'' = \{ V= \{1,4\}, E =\{ (1,4)\} \}$ and we apply $T$. Because $G''_{\pi/4}$ is Clifford we succeed for all outcomes. After the third state, the overall gate implemented is given by $G_{\pi/4}$.}
\end{figure}

\begin{figure}[h]
    \centering
    \subfloat[$T^{(\bo{n})}$]{\includegraphics[width=0.5\columnwidth]{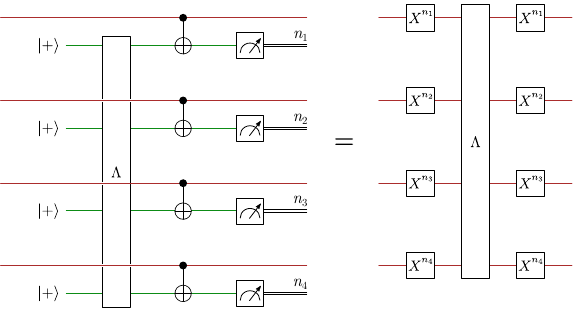} \label{fig.circuit_T}} \\
    \vspace{0.5cm}
    \subfloat[$\widetilde{T}^{(\bo{n},\bo{m})}$]{\includegraphics[width=0.5\columnwidth]{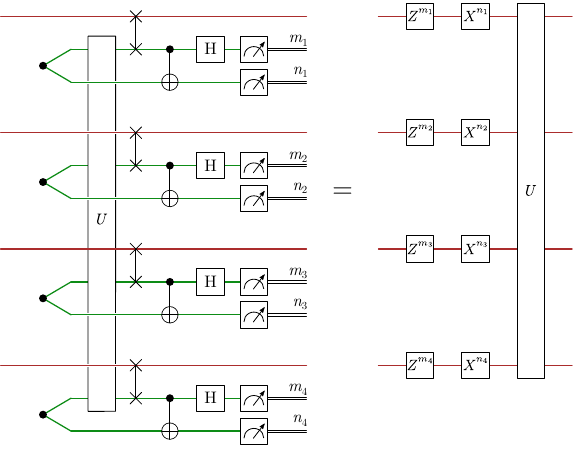} \label{fig.circuit_WT}}   \\
    \vspace{0.5cm}
    \caption{\label{fig:Ualpha} (a) Probabilistic induction of a multiqubit gate $U$. (b) Probabilistic induction of a diagonal gate $\Lambda$. (c) Deterministic implementation of an arbitrary multi-\textit{Z}-rotation with an auxiliary GHZ state.}
\end{figure}

\textbf{3rt step}. We need to implement
\begin{equation}
    G''_{\pi/4} = e^{\ti \frac{\pi}{4} Z_1 Z_4}
\end{equation}
Note that $G''_{\pi/4}$ is Clifford, and therefore it can be deterministically implemented by performing a local correction operation, i.e., if outcome $\bo{k}_3 = (l, \varnothing, \varnothing, k)$ is obtained then the correction operation is given by $(Z_1 Z_4)^{l+k}$, and
\begin{equation}
    (Z_1 \, Z_4)^{l+k} \, X_1^l \, X_4^k \, G''_{\pi/4} \, X_1^l \, X_4^k = G''_{\pi/4}.
\end{equation}
Therefore, after the third step, the overall gate is given by
\begin{equation}
    G''_{\pi/4} \, e^{\ti \frac{3\pi}{8} Z_1 Z_4} e^{\ti \frac{\pi}{16} (Z_1 Z_3 + Z_2 Z_4 + Z_3 Z_4)} = G_{\pi/16}.
\end{equation}
and the implementation is over. See Fig.~\ref{fig:loko}.

\renewcommand{\thesubfigure}{\arabic{subfigure}}

\section{Inducing the Toffoli gate with routine \textit{T}}
\label{Appendix:toffoli}

We show in detail how to induce the Toffoli gate by applying routine $T$. The Toffoloi gate is LU equivalent to
\begin{equation}
    F = e^{\ti \pi \proj{\bo{0}}}.
\end{equation}

\textbf{1st step}: We try to induce $F$ by preparing $\ket{F}_M$ and applying $T$, i.e.,
\begin{equation}
    T_{\textit{MD}}^{(\bo{k}_1)} \! : \, \ket{F}_M \ket{\psi_0}_D \; \mapsto \; \ket{\psi_1}_D = W_1 \ket{\psi_0}_D
\end{equation}
where $W_1 = X^{\otimes \bo{k}_1} F X^{\otimes \bo{k}_1} = e^{\ti \pi \proj{\bo{k}_1}}$. With probability $p_1 = 2^{-n}$, we obtain $\bo{k}_1=\bo{0}$ and $W_1 = F$ is successfully implemented, otherwise, we fail and proceed with step 2.

\textbf{2nd step}: We try to undue $W_1$ and implement $F$, so we prepare $\ket{F W_1}$ (becouse $W_1^\dagger = W_1$) and apply $T$, i.e.,
\begin{equation}
    T_{\textit{MD}}^{(\bo{k}_2)} \! : \, \ket{W_1 F}_M \ket{\psi_1}_D \; \mapsto \; \ket{\psi_2}_D = W_2 \ket{\psi_1}_D = W_2 \, W_1 \ket{\psi_0}_D ,
\end{equation}
$W_2 = X^{\otimes \bo{k}_2} W_1 F X^{\otimes \bo{k}_2} = e^{\ti \pi \left(\proj{\bo{k}_1 \oplus \bo{k_2}} + \proj{\bo{k}_2} \right)}$,
and therefore the gate implemented after the 2nd step is given by 
\begin{equation}
    W_2 \, W_1 = e^{\ti \pi \left(\proj{\bo{k}_1} + \proj{\bo{k}_2} + \proj{\bo{k}_1 \oplus \bo{k_2}} \right)} .
\end{equation}
Note that $W_2 \, W_1 = F$ for $\bo{k}_2 \in \{ \bo{k}_1 , \bo{0} \}$, and therefore, the success probability is $p_2 = 2 \cdot 2^{-n}$. In case of failure, we go with step three.

\textbf{3rd step}: We try to undue $W_2 W_1$ and implement $F$, so we prepare $\ket{F W_1 W_2}$ (becouse $W_2^\dagger = W_2$) and apply $T$, i.e.,
\begin{equation}
    T_{\textit{MD}}^{(\bo{k}_3)} \! :\, \ket{W_2 W_1 F}_M \ket{\psi_2}_D \; \mapsto \; \ket{\psi_3}_D = W_3 \ket{\psi_2}_D = W_3 \, W_2 \, W_1 \ket{\psi_0}_D ,
\end{equation}
where
\begin{equation}
    W_3 = X^{\otimes \bo{k}_3} F \, W_1 \, W_2 \, X^{\otimes \bo{k}_3},
\end{equation}
and hence the gate implemented after the step 3 is given by
\begin{equation}
    W_3 \, W_2 \, W_3 = F \prod_{\bo{k} \in \Xi_3} e^{\ti \pi \proj{\bo{k}}} ,
\end{equation}
where $\Xi_3 = \{ \bo{0}, \bo{k}_1, \bo{k}_2, \bo{k}_3, \bo{k}_1 \oplus \bo{k}_2, \bo{k}_1 \oplus \bo{k}_3, \bo{k}_2 \oplus \bo{k}_3, \bo{k}_1 \oplus \bo{k}_2 \oplus \bo{k}_3 \}$. Note that $W_3 W_2 W_1 = F$ for $\bo{k}_3 \in \Xi_2 = \{ \bo{k}_1, \bo{k}_2 , \bo{k}_1 \oplus \bo{k}_2\}$, and therefore, the success probability is $p_3 = 4 \cdot 2^{-n}$. In case of failure, we go with the 4th step.

\textbf{\textit{j}th step}: We try to undue $W_{j-1} \cdots W_2 \, W_1$ and induce $F$, so we prepare $\ket{F W_1 W_2 \cdots W_{j-1}}$ (becouse $W_{j-1}^\dagger = W_{j-1}$) and apply $T$, i.e.,
\begin{equation}
    T_{\textit{MD}}^{(\bo{k}_j)} \! :\, \ket{W_{j-1} \cdots W_2 W_1 F}_M \ket{\psi_{j-1}}_D \; \mapsto \; \ket{\psi_j}_D = W_j \ket{\psi_{j-1}}_D = W_j \cdots W_2 \, W_1 \ket{\psi_0}_D,
\end{equation}
where
\begin{equation}
    W_j = X^{\bo{k}_j} \, F \, W_{j-1} \cdots W_2 \, W_1 \, X^{\bo{k}_j}
\end{equation}
and the accumulated gate
\begin{equation}
    W_j \cdots W_2 \, W_1 = F \prod_{\bo{k} \in \Xi_j} e^{\ti \pi \proj{\bo{k}}},
\end{equation}
where $\Xi_j = \langle \{ \bo{k}_1, \dots \bo{k}_j \} \rangle_{\oplus}$, and $\langle \{\bullet\}\rangle_{\oplus}$ denotes the set generated by $\{\bullet\}$ with the operation vector sum mod(2) ``$\oplus$''. Note that $\Xi_k$ contains $|\Xi_j| = 2^j$ elements, and we succeed if $\bo{k}_j \in \Xi_{j-1}$. Therefore, the success probability is given by $p_j = |\Xi_{j-1}| \, 2^{-n} = 2^{j-1} \, 2^{-n}$. This means the success probability is doubled in each step. Following the equation $p_j=2^{j-1} \, 2^{-n}$, one would expect that $n+1$ steps are required to achieve a deterministic implementation. However, at the $(n-1)$th step the gate
\begin{equation}
    W_{n-1} \cdots W_1 = F \prod_{\bo{k} \in \Xi_{n-1}} e^{\ti \pi \proj{\bo{k}}},
\end{equation}
is Clifford as we have shown in Sec.~\ref{sec:Diagonal.non-Clifford.gates}, and in case we fail we can correct it with a local operation.

\section{Entanglement cost}
\label{Appendix.Entanglement.cost}

Given a bipartite state $\ket{\psi}_{AB}$, its entropy of entanglement is given by
\begin{equation}
    E = - \text{tr} \left( \rho_A \log_2 \rho_A \right)
\end{equation}
where $\rho_A = \text{tr}_B \proj{\psi}$. The entanglement in the gate state of a two-qubit gate rotation $e^{\ti \theta ZZ}$ is given by
\begin{equation}
    E\!\left( \theta \right) = - x \log_2 (x) - (1-x) \log_2 (1-x)
\end{equation}
where $x = \cos^2(\theta)$.

If we compute the entanglement cost of implementation of a two-qubit $Z$-rotation of the two methods shown in Sec.~\ref{sec:Diagonal.non-Clifford.gates}, we obtain that with the deterministic approach a Bell state is always destroyed and therefore it has a fixed cost of one ebit. On the other hand, sequentially applying routine $T$, we consume $E(2^{k-1} \theta)$ ebits in the $j$th iteration, and therefore the expected value ebits is given by
\begin{equation}
    \left\langle E \, \right\rangle = \sum_{k=1}^\infty \left( \frac{1}{2} \right)^{k-1} E\!\left( 2^{k-1} \theta \right).
\end{equation}
Note that $\left\langle E \, \right\rangle < 1$ for $|\theta|< 0.2245$, which means for these angles it is more efficient to induce $e^{\ti \theta ZZ}$ with the $T$ gate. Otherwise, we would use a bell state.

\renewcommand{\thesubfigure}{\alph{subfigure}}

\end{document}